\documentclass[twocolumn,showpacs,superscriptaddress,amsmath,aps]{revtex4}
\usepackage{graphicx,color}
\usepackage{bm}
\usepackage[hypertex]{hyperref}

\newcommand{\be}{\begin{equation}}
\newcommand{\ee}{\end{equation}}
\newcommand{\bea}{\begin{eqnarray}}
\newcommand{\eea}{\end{eqnarray}}
\newcommand{\bsube}{\begin{subequations}}
\newcommand{\esube}{\end{subequations}}

\newcommand{\Eq}[1]{Eq.\,(\ref{#1})}

\newcommand{\la}{\langle}
\newcommand{\ra}{\rangle}

\newcommand{\nl}{\nonumber \\}

\newcommand{\bsub}{\begin{subequations}}
\newcommand{\esub}{\end{subequations}}

\begin{document}

\title{Direct measurement of the quantum state of photons in a cavity}

\author{Lupei Qin}
\affiliation{Center for Joint Quantum Studies and
Department of Physics, Tianjin University, Tianjin 300072, China}
%
\author{Zhong Wang}
\affiliation{Department of Physics, Beijing Normal University,
Beijing 100875, China}

\author{Cheng Zhang}
\affiliation{Department of Physics, Beijing Normal University,
Beijing 100875, China}

\author{Xin-Qi Li}
\email{lixinqi@bnu.edu.cn}
\affiliation{Center for Joint Quantum Studies and
Department of Physics, Tianjin University, Tianjin 300072, China}
\affiliation{Department of Physics, Beijing Normal University,
Beijing 100875, China}

\date{\today}

\begin{abstract}
We propose a scheme to measure
the quantum state of photons in a cavity.
The proposal is based on the concept of quantum weak values
and applies equally well to both the solid-state circuit
and atomic cavity quantum electrodynamics (QED) systems.
The proposed scheme allows us to access {\it directly}
the superposition components in Fock state basis,
rather than the Wigner function as usual in phase space.
Moreover, the separate access feature held in the direct scheme
does not require a global reconstruction for the quantum state,
which provides a particular advantage beyond
the conventional method of quantum state tomography.
\end{abstract}


\maketitle

{\flushleft The state of a system}
in quantum theory is described by a quantum wavefunction,
which differs drastically from the state description in classical mechanics.
Actually the wavefunction represents a knowledge
and works perfectly well as a practical tool,
however, the underlying physics remains still unclear.
The most surprising point is that the quantum state
is governed by the simple Schr\"odinger equation as a universal law.
Actually, controllable manipulation of the quantum state has
stimulated the advent of the quantum information science and technology.

In addition to manipulating the quantum state
based on the law of Schr\"odinger equation, another important problem
is how to determine a unknown state.
In general, this is a challenging task, since the quantum state
can be determined only by multiple measurements on an ensemble
of identically prepared quantum systems,
rather than a single shot measurement of the single system.
More specifically, to reconstruct the quantum state uniquely,
a complete set of probability distributions
has to be measured over a range of different representations,
by employing the technique of quantum state tomography (QST)
\cite{Ris89,Bre97,Kwi99,Hof09}.

For low dimensional states such as the one of a qubit,
the task is relatively simple. But for high dimensional states,
the job is nontrivial and quite difficult in general.
Particular examples include the determination
of the optical fields in a cavity
\cite{Wil91,Smi93,Fre94,Dut94,Bre95,Bar9596,Ber02,Dav97,Dav01,Seme06}
and of traveling light
\cite{Ban99,Muk03,Ban05,All09a,All09b,Lai10},
the vibrational states of trapped ions/atoms
\cite{Ris92,CZ94,Vog95,CZ96,Mil96,Bar96,Wine96,Ste05}
and molecules \cite{Muk95}.
In these QST schemes for measuring
either the optical fields or the vibrational states,
the strategy is to `measure' the Wigner function
(but not the wavefunction or density operator of state),
by converting the information of the Wigner function
into electronic states of atoms
and performing fluorescence measurement of the atoms.
Viewing that the Wigner function is a class of distributions
in phase space, the uncertainty principle forbids to interpret
it as real probability distribution \cite{Lvo09}.
In order to convert it to {\it real} physical density matrix,
one needs in principle its full information over the phase space,
when performing the transformation
from the Wigner function to quantum density matrix.
This is a demanding task, which requires measuring the Wigner function
over a large grid of points in the phase space.

In this work, we propose a scheme to measure {\it directly}
the quantum wavefunction (but not the Wigner function)
of the optical field (photons) in a cavity,
first in the solid-state circuit QED
then in an atomic cavity QED systems.
Importantly, the proposed scheme allows us to access the individual
superposition component in Fock state basis, and does not
need global reconstruction as usual in the conventional QST scheme.
The new scheme is essentially based on the concept
of quantum weak values (WVs) \cite{Aha88,Ste89,Aha90}.

Actually, the concept of quantum WVs has been exploited for applications
such as `direct' measurement of quantum wavefunctions
\cite{Lun11,Lun12,Lun16,Boy13,Boy14,Boy14a}.
The basic idea is sequentially
measuring two complementary variables of the system.
The first measurement is weak, and the second one is strong.
The weak measurement gets minor information, which
has gentle disturbance and does not collapse the state.
The second projective measurement plays a role of post-selection.
One of the most desirable features is that, in this new scheme,
it is the superposed {\it complex amplitudes} in the wavefunction
(but not the {\it probabilities})
to be extracted from the single round average
of the post-selected data of the first weak measurements.
Another advantage of the WV-based scheme is the possibility
that it does not necessarily
need a global reconstruction of the quantum state.
Applying this method, experiments have been performed
for measuring photon's transverse wavefunction
(a task not previously realized by any method) \cite{Lun11},
photon's polarization state \cite{Lun16,Boy13},
and the high-dimensional orbital
angular momentum state of photons \cite{Boy14,Boy14a}.

\begin{figure}[htb!]
\center
\includegraphics[scale=0.28] {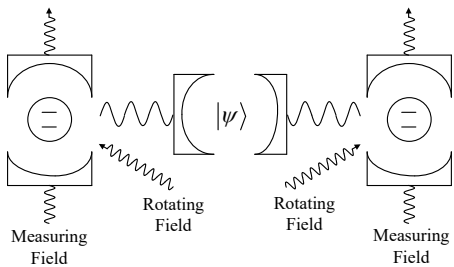} \\
\caption{
Schematic plot for
measuring the unknown state of photons in a cavity,
say, in the central one which can be
expressed in general as $|\Psi\ra=\sum_n c_n |n\ra$
with $|n\ra$ the Fock state of $n$ photons.
In connection with the superconducting circuit-QED realization,
the two artificial atoms (qubits) in the side cavities
are employed to probe the photons state in the central cavity:
the left qubit performs weak measurement selectively
for $\Pi_n=|n\ra \la n|$;
and the right qubit performs post-selection
which will result in a post-selected cavity state
$|\Psi_f\ra=\sum_n c_n (\alpha_n|n\ra + \beta_n|n-1\ra)$.
The coupling between the cavities, the required rotations of qubits
and their measurements are also schematically indicated,
while keeping more detailed explanations referred to the main text. }
\end{figure}

{\flushleft\it Set-up description and basic idea}.---
In Fig.\ 1 we show schematically the proposed set-up
which can be realized with superconducting circuit QED architectures
\cite{Bla04,Wall04,Dev13,Sid13,Mo15}.
The high-Q cavity in the middle part is prepared
in a quantum state of microwave field to be measured.
Taking the most natural choice of representation basis,
the cavity field state can be expressed as
$|\Psi\ra=\sum_n c_n |n\ra$,
where $|n\ra$ is the Fock state with $n$ photons.
The left and right artificial atoms correspond to the transmon qubits
in the circuit QED realization, each of them being stored in its own cavity.
The two qubits are designed to couple to the middle cavity
to jointly probe the cavity field.
More specifically, the left-side (meter) qubit performs weak measurement selectively
for $\Pi_n=|n\ra \la n|$ (with ``$n$" a running number),
and the right-side (post-selection) qubit generates post-selection to the cavity field.
In order to realize the selective monitoring of $\Pi_n$,
the left-side qubit is dispersively coupled to the middle cavity
and the weak interaction with $\Pi_n$
is implemented by performing, e.g., a $\sigma_{1x}$ rotation to the left qubit
by a small angle, by applying a rotating field with frequency
in resonance with the $n$-photon-shifted qubit energy.
Then, perform projective measurements of $\sigma_{1x}$ and $\sigma_{1y}$,
respectively, for the left qubit (in ensemble of realizations), via the well
established technique of microwave transmission and homodyne detection.
Meanwhile, to perform post-selection, the right-side qubit
is time-controllably coupled to the middle cavity.
Rather than dispersive,
here a resonant coupling is proposed.
Together with proper rotation to the qubit and homodyne detection of
microwave transmission (to projectively measure the qubit state),
desired post-selection for the middle cavity state can be realized.
Conditioned on the post-selection, the conditional averages of
$\sigma_{1x}$ and $\sigma_{1y}$ of the left-side qubit
will reveal essential information of the $n_{\rm th}$ component $c_n$
for the quantum state of the middle cavity.    \\
\\
{\it Weak-value and state determination.}---
Now we present more detailed description for the method
how to measure first the weak value of $\Pi_n$,
then determine the unknown state of the cavity field.
As briefly mentioned above, the left-side qubit in Fig.\ 1
is dispersively coupled to the middle cavity, described by
$H_{int}=\chi a^{\dag}a\sigma_{1z}$,
where $a^{\dagger}$ and $a$ are the creation and annihilation
operators of the single mode cavity photons,
$\sigma_{1z}$ is the quasi-spin operator of the left qubit
with logic states $|e_1\ra$ and $|g_1\ra$
(another two operators of this qubit are $\sigma_{1x}$ and $\sigma_{1y}$).
The bare energy spacing between $|e_1\ra$ and $|g_1\ra$ is $2\Delta_1$.
As a consequence of ac-Stark effect
(or, directly, based on the above dispersive Hamiltonian),
the qubit energy will be shifted
from $\Delta_1$ to $\tilde{\Delta}_1^{(n)}=\Delta_1+n\chi$
by the Fock state $|n\ra$ of the cavity field.

In order to realize the measurement of $\Pi_n=|n\ra\la n|$,
let us consider a `selective' $\sigma_{1x}$ rotation on the qubit,
by applying an external microwave field with frequency
in resonance with $2\tilde{\Delta}_1^{(n)}$.
This induces a measurement coupling between the cavity field
and the qubit given by
\begin{equation}\label{H-m1}
    H_{meas}=\gamma \Pi_{n}\sigma_{1x}  \, .
\end{equation}
In this measurement interaction Hamiltonian,
$\gamma$ is the rotating strength to the qubit,
and the projection operator $\Pi_{n}=|n\rangle\langle n|$
is from the fact that we selectively rotate the qubit with frequency
in resonance with $2\tilde{\Delta}_1^{(n)}$.
More quantitative derivation for \Eq{H-m1}
is referred to a latter part in this work.

Under the action of the Hamiltonian \Eq{H-m1}, the cavity field
and the meter qubit (i.e. the left one in Fig.\ 1)
are subject to a joint evolution.
Let us denote the initial state as
$|\Psi\rangle \otimes |\Phi_{0}\rangle$,
where $|\Phi_{0}\rangle$ is the state of the meter qubit
before switching on the measurement interaction,
which is assumed as $|\Phi_{0}\rangle=|g_1\rangle$.
The joint evolution is given by
$U(\tau)(|\Psi\rangle \otimes |\Phi_{0}\rangle)$, where
$U(\tau)=\exp(-iH_{meas} \tau/\hbar)
\simeq 1-i(\frac{\gamma\tau}{\hbar})\Pi_{n}\sigma_{1x}$
in the regime of weak measurement which
is characterized by a small parameter of $\gamma\tau$.
Conditioned on a post-selection of the cavity field state $|\Psi_f\ra$,
which is to be specified soon in the following,
the state of the meter qubit is given by
\begin{eqnarray}\label{Phi-tau}
|\Phi(\tau)\rangle =
\left[ |g_1\rangle-i(\frac{\gamma\tau}{\hbar})
\la\Pi_n\ra_w |e_1\rangle \right]/{\cal N}  \,,
\end{eqnarray}
where ${\cal N}$ denotes a normalization factor
and the weak value $\la\Pi_n\ra_w$ reads
\bea\label{Pi-WV}
\la\Pi_n\ra_w=\frac{\langle \Psi_{f}|\Pi_{n}|\Psi\rangle}
{\langle \Psi_{f}|\Psi\rangle} \,.
\eea
Importantly, the weak value of $\Pi_n$ in \Eq{Phi-tau}
plays a role of rotation parameter to the meter qubit.
Using standard method, this complex parameter can be extracted from the
averages of the meter qubit,
$\la \sigma_{1x}\ra_{\Phi}=\la \Phi(\tau)|\sigma_{1x} | \Phi(\tau)\ra$
and $\la \sigma_{1y}\ra_{\Phi}=\la \Phi(\tau)|\sigma_{1y} | \Phi(\tau)\ra$.
After simple algebra, we obtain
\bea
\langle\Pi_{n}\rangle_{w}
=\left(\frac{\hbar}{2\gamma\tau}\right)
  (i\langle \sigma_{1x}\rangle_{\Phi}
    -\langle \sigma_{1y}\rangle_{\Phi} )  \,.
\eea

The averages $\la \sigma_{1x}\ra_{\Phi}$ and $\la \sigma_{1y}\ra_{\Phi}$
can be obtained via an ensemble of projective measurements
within the `natural' basis $|e_1\ra$ and $|g_1\ra$ of the qubit.
However, before the projective measurements,
a respective $\sigma_{1x}$ or $\sigma_{1y}$ rotation (basis rotation)
should be exerted on the qubit.
Another point associated with the weak value $\langle\Pi_{n}\rangle_{w}$
is that the measurement records are collected only if
the post-selection of the cavity state $|\Psi_f\ra$ is successful.
In our proposal, the average success probability of post-selection
is about 50\%, which is high
among the various weak-value-related applications.

Now we address the post-selection for the cavity field state,
via a couple of procedures in order as follows.
{\it (i)}
Switch on for a time period of
resonant coupling between the cavity field and
the `post-selection' qubit (the right one in Fig.\ 1).
We assume this qubit prepared initially in the ground state $|g_2\ra$.
The coupling interaction leads to a Rabi rotation:
$|g_2\ra|n\ra\rightarrow \alpha_{n}|g_2\ra|n\rangle
+ \beta_{n}|e_2\ra|n-1\rangle$.
{\it (ii)}
Perform, for instance, $\pi/2$-pulse $\sigma_{2y}$ rotation to the qubit,
which is described by the unitary transformation
$U(\theta)=e^{-i\frac{\theta}{2}\sigma_{2y}}$.
After {\it (i)} and {\it (ii)},
the joint state of the cavity and qubit reads:
\bea
|\Psi_2\ra &=&
\frac{1}{\sqrt{2}} \sum^{\infty}_{n=0}c_{n}\,
[\,(\alpha_{n}|n\rangle-\beta_{n}|n-1\rangle)|g_2\rangle  \nl
&& ~~~~~~~~~~
+ (\alpha_{n}|n\rangle+\beta_{n}|n-1\rangle)|e_2\rangle\, ] \,.
\eea
{\it (iii)}
Perform a projective measurement on the qubit and select the result of $|g_2\ra$
we obtain the cavity state as
\bea
|\Psi_{f}\rangle = \left[ c_{0}\alpha_0|0\rangle +  \sum^{\infty}_{n=1} c_{n}
(\alpha_{n}|n\rangle-\beta_{n}|n-1\rangle) \right]/{\cal N} \,.
\eea
This is the post-selected state of the cavity photons.
Inserting it into \Eq{Pi-WV}, up to a common normalization factor, we obtain
\begin{eqnarray}
  \langle \Pi_{0}\rangle_{w} &=& |c_0|^2
  \left[\alpha_0-(c_1/c_0)\beta_1  \right]^*  \,,  \nl
  \langle \Pi_{1}\rangle_{w} &=& |c_1|^2
  \left[\alpha_1-(c_2/c_1)\beta_2  \right]^*  \,,  \nl
   &\vdots&     \nl
\langle \Pi_{n}\rangle_{w} &=& |c_n|^2
  \left[\alpha_n-(c_{n+1}/c_n)\beta_{n+1}  \right]^* \,.
\end{eqnarray}
Up to a common normalization factor,
like other WV-based state tomographic schemes
\cite{Lun11,Lun12,Lun16,Boy13,Boy14,Boy14a},
this set of iterative expressions allows us to determine sequentially
$c_1, c_2, \cdots c_n$,
based on the measured weak values $\la\Pi_n \ra_w$
(note that all the $\alpha_n$ and $\beta_n$ are known coefficients).
Compared to the conventional tomographic method,
which cannot access the individual components of the superposed state,
the present iterative expressions hold the advantage
of permitting us to access the single components
without global reconstruction, viewing the fact that it is
the relative ratios of the amplitudes in the quantum superposition
that represent the real information relevant to observable effects.
Actually, in a quantum superposed state,
the ratio of neighboring components is equivalent to
the relative amplitude with respect to a common normalization factor. \\
\\
{\it Alternative set-up of atomic cavity QED system}.---
The direct scheme of state tomography proposed above can be similarly
applied to the state-of-the-art atomic cavity-QED set-up \cite{Har06}.
The basic idea is schematically illustrated in Fig.\ 2.
The high-Q cavity (`$C$') is prepared in a state
described in general by $|\Psi\ra =\sum_n c_n |n\ra$.
This cavity field is probed first by crossing an atom (meter atom)
through it (as shown by the upper panel of Fig.\ 2),
then by a subsequent post-selection atom (the lower panel of Fig.\ 2).
The second low-Q Ramsey cavity (`$R$')
is employed to rotate the crossing atoms
(sequentially, first the meter and then the post-selection atoms)
between $|g_j\ra$ and $|e_j\ra$ ($j=1,2$),
by introducing $\pi/2$ classical Rabi pulses.

\begin{figure}[htb!]
\center
\includegraphics[scale=0.3] {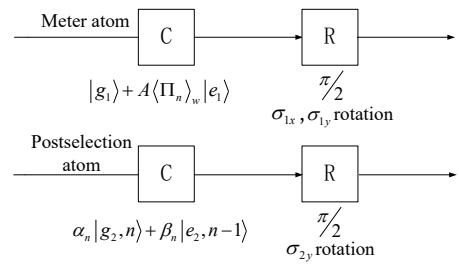} \\
\caption{
Schematic illustration for implementing the proposed scheme
in atomic cavity-QED set-up.
The high-Q cavity (`$C$') is prepared in an initial state
described in general by $|\Psi\ra =\sum_n c_n |n\ra$,
while the second low-Q Ramsey cavity (`$R$')
is employed to rotate the atomic states of
the crossing atoms by introducing external classical fields.
The upper and lower panels show, respectively,
a meter and post-selection atom
crossing sequentially the two cavities. }
\end{figure}

We may detail the weak probe and post-selection of the cavity state,
respectively, as follows.
{\it (i)}
For the meter atom (prepared in ground state $|g_1\ra$ before
entering the cavity $C$), the dispersive coupling with the cavity field
generates an ac Stark shift $n\chi$ between $|g_1\ra$ and $|e_1\ra$,
where $n$ is the photon numbers and $\chi$ the
dispersive coupling strength.
When the meter atom crosses the cavity $C$,
shine a classical laser field into the cavity
to rotate {\it selectively}, i.e., $n$-dependently, the meter atom
weakly by an amount of small angle (small $\gamma\tau$ in Eq.\ (2)).
Then, let the meter atom cross the Ramsey cavity $R$, experience
a $\pi/2$ pulse of $\sigma_{1x}$ and $\sigma_{1y}$ rotations
(in the sense of ensemble realizations),
and suffer a final ionization measurement of $|g_1\ra$ or $|e_1\ra$.
The ensemble averages,
conditioned also on the result (e.g., $|g_2\ra$)
of the subsequent post-selection atom,
give us the key results $\la\sigma_{1x} \ra_{\Phi}$
and $\la\sigma_{1y} \ra_{\Phi}$ required in Eq.\ (4).

{\it (ii)}
In order to generate the post-selection state
$|\Psi_f\ra$ for the cavity field,
a post-selection atom (following the meter atom)
is sending to cross the both cavities $C$ and $R$.
In $C$, this atom experiences a resonant interaction
with the cavity photon; while in $R$, it suffers
a $\pi/2$ Rabi pulse for $\sigma_{2y}$ rotation.
After these, the atom is subject to a final ionization measurement.
Selecting the result of $|g_2\ra$,
we obtain the post-selection state $|\Psi_f\ra$
for the cavity field, given by Eq.\ (6).   \\
\\
{\it On the `selective' rotation.}---
We now present a derivation for the Hamiltonian shown by Eq.\ (1).
Let us return to the starting Hamiltonian
of the meter qubit (the first one) coupling to the cavity mode
and in the presence of driving by external field,
$ H=\Delta_1\,\sigma_{1z} + \chi\sigma_{1z} a^{\dagger}a +
[(\gamma e^{i\omega t})\sigma_1^- + {\rm h.c.}] $,
where the second term describes the dispersive coupling
of the meter qubit to the cavity mode
and the third term is the external driving (with frequency $\omega$).
We may regard the first two terms as free Hamiltonian,
$ H_0=\Delta_1\,\sigma_{1z} + \chi\sigma_{1z} a^{\dagger}a $
and express it as a sum from subspaces expanded by
$\{(|e_1,n\ra,\,|g_1,n\ra\}$ (with $n=0,1,2,\cdots$):
$H_0=\bigoplus_n H^{(n)}_{0}=\bigoplus_n \tilde{\Delta}_1^{(n)} \sigma^{(n)}_{1z}$,
where $\tilde{\Delta}_1^{(n)}=\Delta_1 + n\chi$ and $H_0^{(n)}$ reads
\begin{eqnarray}
H_0^{(n)}=\begin{pmatrix}
\tilde{\Delta}_1^{(n)}  & 0   \\
0   &  - \tilde{\Delta}_1^{(n)}   \end{pmatrix}   \,,
\end{eqnarray}
Now, including the driving term and in the rotating frame
with respect to $\tilde{H}_0=\frac{\omega}{2}\sigma_{1z}$,
we can express the Hamiltonian in the $n_{\rm th}$ subspace as
\begin{eqnarray}
H^{(n)}=\begin{pmatrix}
\tilde{\Delta}_1^{(n)}- \frac{\omega}{2}  & \gamma   \\
\gamma   &  -(\tilde{\Delta}_1^{(n)}- \frac{\omega}{2})  \end{pmatrix}   \,.
\end{eqnarray}
Note that in terms of this decomposition, the total Hamiltonian
simply reads $H=\bigoplus_n H^{(n)}$.

Consider now the initial state,
$|g_1\ra\otimes|\Psi\ra = |g_1\ra\otimes(\sum_n c_n|n\ra)$.
If we choose the frequency of the driving field in resonance with
the shifted energy of the qubit by $n$ photons, i.e., $\omega=2 \tilde{\Delta}_1^{(n)}$,
only the state component in the $n_{\rm th}$ subspace
will be affected by the driving field.
That is, $|g_1,n\ra$ is rotated by a small amount as
\bea
|g_1,n\ra \longrightarrow  |g_1,n\ra - i(\gamma\tau/\hbar) |e_1,n\ra \,.
\eea
Here, we expanded the unitary evolution operator $U(\tau)$
to the first order, which is valid in the weak measurement limit.
Other components in $|g_1\ra\otimes|\Psi\ra$,
owing to large detuning from the frequency of the driving field,
are not affected by the driving field.
Putting these together, we have
\bea
&& U(\tau)\,\left[|g_1\ra\otimes(\sum_n c_n|n\ra)\right] \nl
&=& \sum_{n'\neq n} c_{n'} |g_1,n'\ra
+ c_n [|g_1,n\ra - i(\gamma\tau/\hbar)\, |e_1,n\ra]   \nl
&=& |g_1\ra\otimes|\Psi\ra -i(\gamma\tau/\hbar)\, c_n|e_1,n\ra \nl
&=& |g_1\ra\otimes|\Psi\ra - i (\gamma\tau/\hbar)|e_1\ra\otimes(\hat{\Pi}_n|\Psi\ra)  \,.
\eea
This allows us to construct the {\it effective} rotating Hamiltonian, Eq.\ (1),
which leads to the {\it selective} rotation given by Eq.\ (2).


Finally, let us explain how the state in the subspace with large energy detuning
can be free from the influence of the rotating field.
In the rotating frame with frequency $\omega=2\tilde{\Delta}_1^{(n)}$,
the detuning of the $n'$-photon-shifted qubit energy from $\omega$
is characterized by nonzero energies of the qubit states
$|e_1\ra$ and $|g_1\ra$, $E_{e_1,g_1}=\pm\epsilon$,
where $\epsilon=|\tilde{\Delta}_1^{(n')}-\omega/2|=|n'-n|\chi$.
Then, after a simple algebra, the transition probability
from $|g_1\ra$ to $|e_1\ra$ is obtained as
\bea
P_{e_1}(t)=\left(\gamma/\tilde{\gamma} \right)^2
\sin^2(\tilde{\gamma}t) \,,
\eea
where $\tilde{\gamma}=\sqrt{\epsilon^2+\gamma^2}$.
In the special case of resonant driving (i.e. $\epsilon=0$)
and for weak measurement limit, we have
\bea\label{Pe-1}
P_{e_1}(t) = \sin^2(\gamma t) \simeq (\gamma t)^2  \,.
\eea
For nonzero energy detuning, we reexpress the transition probability as
\bea\label{Pe-2}
P_{e_1}(t)=(\gamma t)^2 \left[\frac{\sin^2(\tilde{\gamma}t)}
{(\tilde{\gamma}t)^2}   \right]  \,.
\eea
Let us assume that the weak measurement transition given by the upper result \Eq{Pe-1}
is realized by weak coupling (with small $\gamma$).
Then, under the condition of strong dispersive coupling $\chi/\gamma>>1$,
the $\tilde{\gamma}$ in the lower result \Eq{Pe-2} can be approximated as
$\tilde{\gamma}\simeq \epsilon=|n'-n|\chi\equiv m\chi$.
Now, importantly, if we properly design the coupling strength and time
to make $\gamma t$ a small parameter and $\chi t\simeq \pi$,
based on \Eq{Pe-2} we find that,
for the $n'(\neq n)$-photon-shifted qubit state,
the transition from $|g_1\ra$ to $|e_1\ra$ is to be strongly suppressed
owing to $\sin^2(m\chi t)/(m\chi t)^2\to 0$.
Therefore, via this type of design,
we can realize the `selective' rotation of the $n$-photon-shifted state. \\
\\
{\it Discussion and Summary.}---
One of the subtle issues in practice
is the accurate reset of the initial state of cavity field,
after each weak measurement and postselection.
This is because the second postselection would destroy the cavity photons state,
despite the negligible influence on it of the first weak measurement.
The reset can be fulfilled by properly driving the cavity by external field,
and/or coupling it to qubits (e.g. in the solid-state circuit QED architecture),
or sending a stream of atoms to cross through the cavity to excite
cavity photons (e.g. in the case of atomic cavity QED set-up).
Apparently, the accuracy of the reset will set up the upper limit of
tomography quality, as in any other tomographic schemes,
owing to the probability nature of the quantum wavefunction.

Other issues in experiment include properly performing both
the $\sigma_x$ and $\sigma_y$ rotations -- this can be realized
by modulating the phase of the driving field by $\pi/2$,
and precisely tuning the selective frequency of the weak measurement
in resonance with $\Delta_1+n\chi$.
This frequency tuning can be implemented by (i) altering the
frequency of the driving field,
and/or (ii) modulating the level spacing of the qubit
by gate voltage control (in the case of circuit QED set-up).

Existing tomographic schemes of cavity field is
measuring the Wigner function in phase space.
In order to convert the Wigner function to
density matrix in physical state representation,
one needs to digitalize the phase space and
gain by measurement the database of a large grid of points.
For each of these points, one must perform the usual ensemble measurements.
In contrast, the present WV-based scheme provides a direct access
to the individual Fock-state component we desired of the cavity field,
not needing a global reconstruction of the whole quantum state.
Actually, we may understand the present scheme
is a {\it conjugated} one of the Wigner function measurement.
In concern with the extra procedure of postselection involved
in the WV-based scheme,
our present proposal holds a feature of high efficiency,
viewing that the postselection of the cavity field
is fulfilled by selecting one from the two states of the second qubit/atom,
which is also unaffected by the average photon number
of the cavity field subject to measurement.
This high efficiency postselection can benefit a lot to the practical
realization of the present scheme,
by regarding the high dimensions of the cavity photons state.

To summarize, we have proposed a scheme
to measure the quantum state of photons in a cavity.
The scheme is essentially based on the concept
of quantum weak values, which allows direct access to
the individual superposition components in Fock state basis,
not needing a global reconstruction as the
conventional method of quantum state tomography.
Compared to existing schemes of measurement of
the Wigner function, the present scheme does not need
the conversion from phase space to physical representation.
It would be of particular interest to realize the proposal
in the state-of-the-art superconducting circuits.

\vspace{0.2cm}
{\flushleft \it Acknowledgments.}---
This work was supported by
the NNSF of China under Nos. 11675016 \& 21421003.



\begin{references}
\bibitem{Ris89}
K. Vogel and H. Risken, Phys. Rev. A {\bf 40}, 2847 (1989).
\bibitem{Bre97}
G. Breitenbach, S. Schiller, and J. Mlynek,
Nature {\bf 387}, 471 (1997).
\bibitem{Kwi99}
A. G. White, D. F. V. James, P. H. Eberhard,  and P. G. Kwiat,
Phys. Rev. Lett. {\bf 83}, 3103 (1999).
\bibitem{Hof09}
M. Hofheinz {\it et al.}, Nature {\bf 459}, 546 (2009).


\bibitem{Wil91}
M. Wilkens and P. Meystre, Phys. Rev. A {\bf 43}, 3832 (1991).
\bibitem{Smi93}
D. T. Smithey, M. Beck, M. G. Raymer, and A. Faridani,
Phys. Rev. Lett. {\bf 70}, 1244 (1993).
\bibitem{Fre94}
M. Freyberger and A. M. Herkommer, Phys. Rev. Lett. {\bf 72}, 1952 (1994).
\bibitem{Dut94}
S. M. Dutra and P. L. Knight, Phys. Rev. A {\bf 49}, 1506 (1994).
\bibitem{Bre95}
G. Breitenbach, T. M\"uller, S. F. Pereira, J. Ph. Poizat,
S. Schiller, and J. Mlynek, J. Opt. Soc. B {\bf 12}, 2304 (1995).
\bibitem{Bar9596}
P. J. Bardroff {\it et al.}, Phys. Rev. A {\bf 51}, 4963 (1995);
{\bf 53}, 2736 (1996).
\bibitem{Ber02}
P. Bertet, A. Auffeves, P. Maioli, S. Osnaghi, T. Meunier, M. Brune,
J.M. Raimond, and S. Haroche, Phys. Rev. Lett. {\bf 89}, 200402 (2002).
\bibitem{Dav97}  
L. G. Lutterbach and L. Davidovich, Phys. Rev. Lett. {\bf 78}, 2547 (1997).
\bibitem{Dav01}
M. Franca Santos, L. G. Lutterbach, S. M. Dutra, N. Zagury, and L. Davidovich
Phys. Rev. A {\bf 63}, 033813 (2001).
\bibitem{Seme06}
A. A. Semenov, D. Yu. Vasylyev, W. Vogel, M. Khanbekyan, and D.-G. Welsch
Phys. Rev. A {\bf 74}, 033803 (2006).



\bibitem{Ban99}
K. Banaszek {\it et al.}, Phys. Rev. A {\bf 60}, 674 (1999).
\bibitem{Muk03}
E Mukamel, K Banaszek, and I A. Walmsley, and C Dorrer,
Opt. Lett. {\bf 28}, 1317 (2003);
\bibitem{Ban05}
B J. Smith, B Killett, and M. G. Raymer, I. A. Walmsley,
K. Banaszek, Opt. Lett. {\bf 30}, 3365 (2005).
\bibitem{All09a}
M. Bondani, A. Allevi, and A. Andreoni, Opt. Lett. {\bf 34}, 1444 (2009).
\bibitem{All09b}
A. Allevi {\it et al.}, Phys. Rev. A {\bf 80}, 022114 (2009).
\bibitem{Lai10}
K Laiho, K N Cassemiro, D Gross, and C Silberhorn,
Phys. Rev. Lett. {\bf 105}, 253603 (2010).


\bibitem{Ris92}
C. A. Blockley, D. F. Walls, and H. Risken, Europhys. Lett. {\bf 77}, 509 (1992).
\bibitem{CZ94}
J. I. Cirac, R. Blatt, A. S. Parkins, and P. Zoller,
Phys. Rev. A {\bf 49}, 1202 (1994).
\bibitem{Vog95}
S. Wallentowitz and W. Vogel, Phys. Rev. Lett. {\bf 75}, 2932 (1995).
\bibitem{CZ96}
J. F. Poyatos, R. Walser, J. I. Cirac, P. Zoller, and R. Blatt,
Phys. Rev. A {\bf 53}, R1966 (1996).
\bibitem{Mil96}
C. D'Helon and G. J. Milburn, Phys. Rev. A {\bf 54}, R25 (1996).
\bibitem{Bar96}
P. J. Bardroff et al., Phys. Rev. Lett. {\bf 77}, 2198 (1996).
\bibitem{Wine96}
D. Leibfried， D. M. Meekhof, B. E. King, C. Monroe,
W. M. Itano, and D. J. Wineland, Phys. Rev. Lett. {\bf 77}, 4281 (1996).

\bibitem{Ste05}
J. F. Kanem, S. Maneshi, S. H. Myrskog, and A. M. Steinberg,
J. Opt. B {\bf 7}, S705 (2005).  


\bibitem{Muk95}
T. J. Dunn, I. A. Walmsley, and S. Mukamel,
Phys. Rev. Lett. {\bf 74}, 884 (1995).

\bibitem{Lvo09}
A. I. Lvovsky and M. G. Raymer, Rev. Mod. Phys. {\bf 81}, 299 (2009).


\bibitem{Aha88}
Y. Aharonov, D. Albert, and L. Vaidman,
Phys. Rev. Lett. {\bf 60}, 1351 (1988).
\bibitem{Ste89}
I. M. Duck, P. M. Stevenson, and E. C. G. Sudarshan,
Phys. Rev. D {\bf 40}, 2112 (1989).
\bibitem{Aha90}
Y. Aharonov and L. Vaidman, Phys. Rev. A {\bf 41}, 11 (1990).


\bibitem{Lun11}
J. S. Lundeen, B. Sutherland, A. Patel, C. Stewart, and C. Bamber,
Nature {\bf 474}, 188 (2011).
\bibitem{Lun12}
J. S. Lundeen and C. Bamber,
Phys. Rev. Lett. {\bf 108}, 70402 (2012).
\bibitem{Lun16}
G. S. Thekkadath, L. Giner, Y. Chalich, M. J. Horton,
J. Banker, and J. S. Lundeen,
Phys. Rev. Lett. {\bf 117}, 120401 (2016).
\bibitem{Boy13}
J. Z. Salvail, M. Agnew, A. S. Johnson,
E. Bolduc, J. Leach, and R. W. Boyd,
Nature Photonics {\bf 7}, 316 (2013).
\bibitem{Boy14}
M. Malik, M. Mirhosseini, M. P. J. Lavery, J. Leach,
M. J. Padgett, and R. W. Boyd,
Nature Communications {\bf 5}, 3115 (2014).
\bibitem{Boy14a}
M. Malik and R. W. Boyd,
{\it Quantum Imaging Technologies}, arXiv:1406.1685;
Rivista del Nuovo Cimento 37, 5 (2014) p. 273


\bibitem{Bla04}   
A. Blais, R. S. Huang, A. Wallraff, S. M. Girvin,
and R. J. Schoelkopf, Phys. Rev. A {\bf 69}, 062320 (2004).
\bibitem{Wall04}
A. Wallraff, D. I. Schuster, A. Blais, L. Frunzio,
R. S. Huang, J. Majer, S. Kumar, S. M. Girvin,
and R. J. Schoelkopf, Nature {\bf 431}, 162 (2004).
\bibitem{Dev13}
M. Hatridge, S. Shankar, M. Mirrahimi, F. Schackert, K.
Geerlings, T. Brecht, K. M. Sliwa, B. Abdo, L. Frunzio,
S. M. Girvin, R. J. Schoelkopf, and M. H. Devoret,
Science {\bf 339}, 178 (2013).
\bibitem{Sid13}
K. W. Murch, S. J. Weber, C. Macklin and I. Siddiqi,
Nature {\bf 502}, 211 (2013).
\bibitem{Mo15}
D. Tan, S. J. Weber, I. Siddiqi, K. Molmer, and K.W. Murch,
Phys. Rev. Lett. {\bf 114}, 090403 (2015).


\bibitem{Har06}
S. Haroche and J.M. Raimond. {\it Exploring the Quantum:
atoms, cavities and photons}, Oxford University Press, Oxford (2006).



\end{references}
\end{document}